\documentstyle[12pt,epsfig,wrapfig]{article}

\begin{document}

\topmargin -12mm
\oddsidemargin -4mm

\setcounter{page}{1}
\begin{center}
\bf THE POLARIZATION OF RADIATION IN SINGLE CRYSTALS IN THE
QUASICLASSICAL APPROACH \footnote{The work is supported by ISTC grant A-099}\\
\vspace{2mm}
{\bf S.M.Darbinyan$^a$ and N.L.Ter-Isaakyan $^b$}\\
\vspace{2mm}
{\small Yerevan Physics Institute Yerevan 375036, Armenian\\
$^a$ E-mail: simon@lx2.yerphi.am\\
$^b$ E-mail: terisaak@jerewan1.yerphi.am}\\
\end{center}
%\vspace{2mm}
\begin{abstract}
\indent
The radiation emission spectra of polarized photons emitted from
charged particles in single oriented crystals are obtained in
Bayer-Katkov quasiclassical approach. The results of
numerical calculations are presented in the region of small
angles of incidence for which the coherent theory fails but
magnetic bremsstrahlung region isn't yet achieved. The spectral
distribution of linear polarization degree repeats, in general,
the form of the intensity distribution. At sufficiently small
angles, in the case of planar orientation of crystal, there is a
wide maximum of essential linear polarization at low frequencies
$\omega/\varepsilon=0.1-0.3$. \\
\end{abstract}
\indent
The processes of photon emission from charged particles and
$e^{-}e^{+}$ pair creation by photon at high energies in oriented
single crystals are widely applied in experimental physics for
the production of polarized photon beams, as well as for the
analyses of photon polarization. For sufficiently large incident angles
$\vartheta_{0}$ to crystal axis/planes $\vartheta_{0}\gg\vartheta_{v}$, where
$\vartheta_{v}$ is the characteristic angle given by
$\vartheta_{v}=U_{0}/m$
($U_{0}$ is the scale of axial/planar potential and $m$ the electron mass)
 these processes are well
described in the theory of coherent bremsstrahlung (CB) and pair       
production (CPP) [1]. This theory is constructed in the framework of the
first Born approximation in crystal potential and fails at small angles   
and very high energies, where these processes become of magnetic
bremsstrahlung nature.\\
\indent
 Recently V.N. Bayer with co-authors developed a
general theory for radiation emission from high-energy particles and pair 
production by high-energy photons in strong crystalline fields [2]. This
 theory is not restricted by the first Born approximation and is
 based on the quasiclassical character of motion of ultrarelativistic
particles in strong fields. For  $\vartheta_{0}\gg\vartheta_{v}$,
this theory is reduced to the
standard theory for CB and CPP. At very small angles
$\vartheta_{0}\ll\vartheta_{v}$,  the
quasiclassical theory reproduces the results of constant field
approximation (CFA). For intermediate angles,
$\vartheta_{0}\sim\vartheta_{v}$,
the numerical
calculations in the exact theory face serious mathematical difficulties
and to obtain  specific numerical results some approximation and modeling
methods have been developed.  First numerical results of the radiation
emission spectra in the quasiclassical theory were published in [3] (in 
the framework of additional modeling assumption); exact calculations were
presented in [4]. In Ref. [5] an analytical method of calculations in the
framework of quasiclassical approach was developed and new numerical      
results were presented. These papers demonstrate an essential difference
of exact spectra from the corresponding results of coherent theory at 
small angles $\vartheta_{0}\le\vartheta_{v}$.
There is  good agreement between first experimental
results at small angles [6,7] and these calculations.\\
\indent
Polarization characteristics of radiation emission in single crystals 
in the region of
applicability of CB are well investigated in experiments (at the energies
of incident electrons up to 10 GeV) and the CB theory describes
satisfactory the experimental results [1].\\
\indent
It is very important to
investigate the polarization phenomena at very high energies in oriented
crystals, where the CB theory doesn't work. If we could find optimal
crystal orientations, which bring to the large intensity enhancements of
essentially polarized photons at large frequencies, it will give us an
interesting possibility of producing  high energy polarized photon
beams. However, at energies of 100 GeV and higher the polarization
characteristics of radiation emission have not yet been investigated
experimentally and there are no publications presenting numerical
calculations in the quasiclassical theory.\\
\indent
In this paper the formulae for
radiation emission spectra of polarized photons are derived in
Bayer-Katkov quasiclassical approach. We also present preliminary results 
of numerical calculations in the most interesting region of
intermediate angles $\vartheta_{0}\le\vartheta_{v}$ for planar orientation
of crystal, where maximal polarization effects are expected.To carry out 
the arising integrals over time of complicated oscillating functions
we have elaborated the calculation algorithm and the special integration
program, which is very effective especially at small angles.\\
\indent
  We start from the general Bayer-Katkov quasiclassical formula [2]
which gives the spectral distribution of polarized photon energy, averaged
over the initial and summed over the final electron polarizations:\\  
\begin{equation}
\label{AB}
dE=\alpha\frac{d^{3}k}{(2\pi)^{2}}\int_{-\infty}^{\infty}\,dt_{1}
\int_{-\infty}^{\infty}\,dt_{2}\frac{1}{4\varepsilon'^{2}}[
(\varepsilon+\varepsilon')^{2}({\bf{ev}}_{2})
({\bf{e}}^{*}{\bf{v}}_{1})
-\omega^{2}({\bf{e}}
{\bf{v}}_{1})({\bf{e}}^{*}{\bf{v}}_{2})\\
\end{equation}
$$+\omega^{2}({\bf{v}}_{1}{\bf{v}}_{2}-1+\frac{1}{\gamma
^{2}})({\bf{ee}}^{*})]
e^{ik'(x_{1}-x_{2})}$$

\noindent where $\alpha=1/137$ is fine structure constant,
$k^{\mu}=(\omega,\bf{k})$ and $\bf{e}$- stands,
correspondingly, for photon 4-momenta and polarization vector;
$\varepsilon,\varepsilon'=\varepsilon-\omega$ -
stands for initial and final electron energies,
$k'^{\mu}=k^{\mu}\varepsilon/\varepsilon'\mbox{ , }
\gamma=\varepsilon/m\mbox{ ; }
{\bf{r}}_{1,2}={\bf{r}}(t_{1,2}) \mbox{ and }
{\bf{v}}_{1,2}={\bf{v}}(t_{1,2})$
are electron coordinates and velocities at the moment of time
$t_{1,2}\mbox{ , }x_{1,2}=(t_{1,2},{\bf{r}}_{1,2})$ .\\
\indent
We are going to find the radiation emission spectra of polarized
photons, integrated over the emitted photon angles $\vartheta,\varphi$. 
Therefore, the photon polarization vectors must be defined via the direction 
which remains fixed (keeping in mind smallness of $\vartheta\mbox{ , }
\vartheta\le1/\gamma$ ) after
 integration over $\vartheta,\varphi$ . We chose two independent
polarization vectors in the following form:\\
\begin{equation}
\label{AC}
{\bf{e}}_{1}=\frac{[{\bf{n}}_{2}{\bf{n}}]}
{\left|[{\bf{n}}_{2}{\bf{n}}]\right|}\mbox{ , }
{\bf{e}}_{2}=[{\bf{ne}}_{1}]
\mbox{ , }
\end{equation}
where ${\bf{n}}={\bf{k}}/\omega\mbox{ , }{\bf{n}}_{2}$ is
a unit vector transverse to the incident electron velocity
${\bf{v}}_{0}$,
the exact directions of ${\bf{n}}_{2}$ can be fixed by experimental layout.
We choose ${\bf{n}}_{2}$  transverse to the crystal axis, along which the
incident electron
is aligned.  At $\vartheta\ll1$ polarization vectors ${\bf{e}}_{1}$
and ${\bf{e}_{2}}$ have the same fixed
directions for any $\varphi$  , and the Stocks parameters defined in this
vectors, correspond to the
definite directions after integration over emitted
photon angles.\\
\indent
Utilizing the method derived in ref.[2], we integrate (1) over
photon angles $\vartheta,\varphi$ and go over to the intensity $I=dE/dt$.
If longitudinal length of 
crystal is much higher than radiation formation length, after averaging
over electron trajectories in crystal, we can neglect the intensity
dependence on time [2]. Finally, we present the polarized photon emission 
spectra in terms of Stocks parameters
$\xi_{1}$, $\xi_{2}$, $\xi_{3}$:\\
\begin{equation}
\label{AD}
\frac{dI}{d\omega}=\frac{\alpha}{\pi}\frac{m^{2}\omega}{\varepsilon^{2}}
\int\frac{d^{3}r_{0}}{V} F({\bf{r_{0}}},\vartheta_{0})\left[
\int_{0}^{\infty}\,\frac{d\tau}{\tau}[(T_{0}+\gamma^{2}\xi_{1}T_{1}
+\gamma^{2}\xi_{3}T_{3})
\sin A_{1}+\gamma^{2}\xi_{2}T_{2}]-\frac{\pi}{2}\right]
\end{equation}
Where
 ${\bf{r}}_{0}$ is the entry point, $V$ is crystal volume,
integration over ${\bf{r}}_{0}$ gives the
averaging over electron trajectories in crystal, 
$F({\bf{r}}_{0},\vartheta_{0})$ is the coordinate
distribution function at given value of $\vartheta_{0}$.  The quantities of
$A_{1}=A_{1}(\tau)$ and $T_{i}=T_{i}(\tau),i=0,1,2,3$, in (3) are
defined as follows:
\begin{equation}
\label{AH}
A_{1}=\frac{m^{2}\omega}{2\varepsilon\varepsilon'}
\left[1+\gamma^{2}\left[\frac{1}{\tau}\int_{-\tau/2}^{\tau/2}
{\bf{v}}^{2}_{\perp}(t)\, dt
-{\bf{a}^{2}}_{\perp}(\tau)\right]\right]
\end{equation}
\begin{equation}
\label{AI}
T_{0}=1+\frac{1}{4}\varphi(\varepsilon)\gamma^{2}
({\bf{v}}_{1\perp}-{\bf{v}}_{2\perp})^{2}
\mbox{ , }
\end{equation}
\begin{equation}
\label{AE}
T_{1}=-2a_{x}a_{y}+a_{y}(v_{1x}+v_{2x})+a_{x}(v_{1y}+v_{2y})
-(v_{1x}v_{2y}+v_{1y}v_{2x})
\end{equation}

\begin{equation}
\label{AF}
T_{2}=\frac{\varphi(\varepsilon)}{2}
({\bf{a}}[{\bf{v}}_{1}{\bf{v}}_{2}])
\end{equation}
\begin{equation}
\label{AJ}
T_{3}=a^{2}_{y}-a^{2}_{x}+a_{x}(v_{1x}+v_{2x})-a_{y}(v_{1y}+v_{2y})-
(v_{1x}v_{2x}-v_{1y}v_{2y})
\end{equation}
\label{AJ}

\noindent where ${\bf{a}}(\tau)=\frac{1}{\tau}\int_{-\tau/2}^{\tau/2}{\bf{v}}(t)dt$,
~$\varphi(\varepsilon)=\frac{\varepsilon'}{\varepsilon}+\frac{\varepsilon}
{\varepsilon'}$, ~${\bf{v}}_{\perp}=(v_{x},v_{y})$, 
${\bf{v}}_{1}={\bf{v}}(-\tau/2)$, ~${\bf{v}}_{2}={\bf{v}}(\tau/2)$.
\vspace{2mm}

\noindent The Stocks parameters and projections of vectors
correspond to the frame defined by following unit vectors:
\begin{equation}
\label{AT}
\hat{e}_{x}=[{\bf{n}}_{2}{\bf{v}}_{0}]\mbox{ ,
}\hat{e}_{y}={\bf{n}}_{2}
\mbox{ , }\hat{e}_{z}={\bf{v}}_{0}\mbox{ , }
\end{equation}
i.e. $v_{x}={\bf{v}}\hat{e}_{x}\mbox{ , }v_{y}={\bf{v}}\hat{e}_{y}$ and so
on.\\
\indent
The formulae for polarized photon emission spectra in
quasiclassical approach were derived earlier in ref.[2] and an example of
numerical calculations (in the range of applicability of modified coherent
theory) was presented in [3]. Our results for unpolarized photon
$(T_{0})$  and  circular polarized photon $(T_{2})$ coincide with [2]. For
linear polarization $(T_{1} \mbox{ and } T_{3})$
 our formulae are completely different and the results of our
numerical calculations (see below) differ from [3] even qualitatively! The
difference is originated in the choice of the photon polarization
vectors. In ref.[2] the independent photon polarization vectors were
chosen, correspondingly, in the reaction plane and transverse to it. At
such a choice, if we are interested in the polarization characteristics
of emitted photon, we should not immediately integrate the polarized
spectra over azimuth angle $\varphi$ , because after such integration we
have no fixed direction perpendicular to incident electron velocity, and 
therefore the Stocks parameters cannot be connected with any fixed direction 
in space. One can see that in such calculations even the quantity 
$P=\sqrt{\xi^{2}_{1}+\xi^{2}_{2}}$ cannot be
associated with the degree of linear polarization of emitted photon. For
instance, the value of P carried out in such a way for amorphous medium
does not vanish after integration over $\varphi$ !\\
\indent
For further calculations we shall use the rectilinear trajectory
approximation which is valid, strongly speaking, for
$\vartheta_{0}\gg\vartheta_{c}$ , where $\vartheta_{c}$ is the
Lindhard channeling angle [8].
Substituting into (4)-(8) the electron transverse velocity in crystal in
rectilinear trajectory approximation, we finally find the following
equations for the quantities $A_{1}\mbox{ , } T_{i}$ which determine
the polarized photon emission spectra of Eq.(3):
\begin{equation}
\label{AL}
A_{1}=\frac{m^{2}\omega\tau}{\varepsilon\varepsilon'}
\left[1+\sum_{\bf{q},\bf{q'}}
\frac{G({\bf{q}})G({\bf{q'}})}
{m^{2}q_{z}q'_{z}}{(\bf{q_{\perp}q'_{\perp}})}
\left[\frac{\sin ((q_{z}+q'_{z})\tau)}{(q_{z}+q'_{z})\tau}-
\frac{\sin (q_{z}\tau)\sin(q'_{z}\tau)}{q_{z}q'_{z}\tau^{2}}
\right]e^{-i({\bf{q}}+{\bf{q'}}){\bf{r}}}\right]
\end{equation}
\begin{equation}
\label{AK}
T_0=\left[ 1-\varphi(\varepsilon)\sum_{\bf{q},\bf{q'}}
\frac{G({\bf{q}})G({\bf{q'}})} {m^{2}q_{z}q_{z'}}
({\bf{q}}_{\perp}{\bf{q}}'_{\perp})
\sin (q_{z}\tau)\sin(q'_{z}\tau)e^{-i({\bf{q}}+{\bf{q}'})
{\bf{r}}}\right]\mbox{,}
\end{equation}
\begin{equation}
\label{AO}
T_{1}=-\sum_{\bf{q},\bf{q}'}
\frac{G({\bf{q}})G({\bf{q}}')}
{m^{2}q_{z}q'_{z}}
(q_{x}q'_{y}+q_{y}q'_{x})
e^{-i({\bf{q}}+{\bf{q}}')
{\bf{r}}}[g(q_{z}\tau)g(q'_{z}\tau)+
\sin (q_{z}\tau)\sin (q'_{z}\tau)]\mbox{ ,}
\end{equation}
\begin{equation}
\label{AM}
T_{2}=\varphi(\varepsilon)\sum_{\bf{q},
\bf{q}'} \frac{G({\bf{q}})G({\bf{q}}')}
{m^{2}q_{z}q'_{z}}(q_{x}q'_{y}-q_{y}q'_{x})
e^{-i({\bf{q}}+{\bf{q}}'){\bf{r}}}
[g(q_{z}\tau)\sin (q'_{z}\tau)-g(q'_{z}\tau)\sin
(q_{z}\tau]\mbox{ ,}
\end{equation}
\begin{equation}
\label{AN}
T_{3}=-\sum_{\bf{q},\bf{q}'}
\frac{G({\bf{q}}),G({\bf{q}}')}{m^{2}q_{z}q'_{z}}
(q_{x}q'_{x}-q_{y}q'_{y})e^{-i({\bf{q}}+{\bf{q}}')
{\bf{r}}}[g(q_{z}\tau)g(q'_{z}\tau)+\sin
(q_{z}\tau)\sin (q'_{z}\tau)]\mbox{ ,}
\end{equation}
$$
g(x)=\frac{\sin x }{x}-\cos x\mbox{ .}
$$
\begin{wrapfigure}{r}{10.cm}
\epsfig{figure=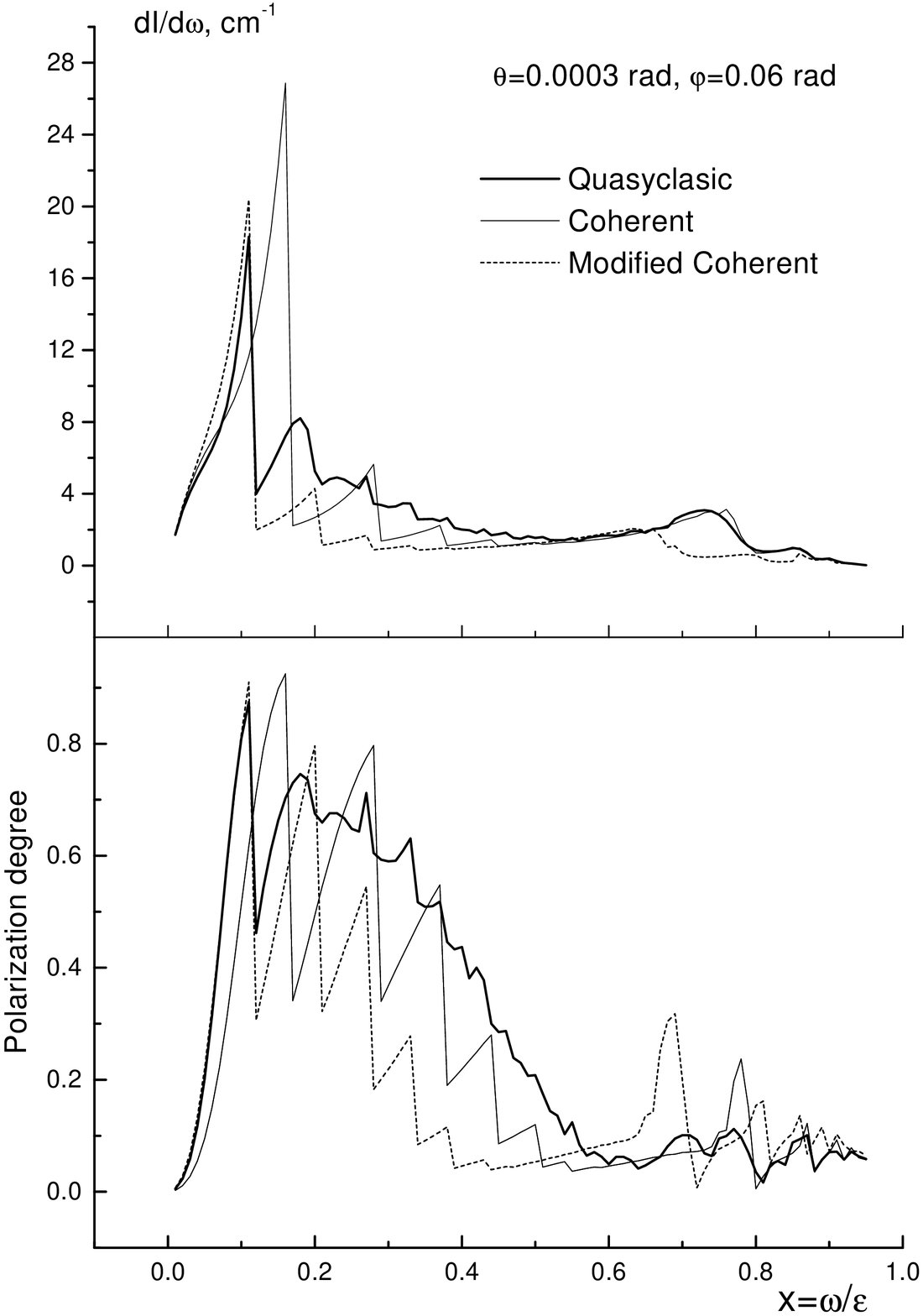,width=10.4cm}
{\small Fig.1: Comparison the spectral distributions of intensity and degree of
polarization in coherent [1] and quasiclassical [2] approaches at
relatively large angles.}
\end{wrapfigure}
\noindent The notation used here coincide, in general, with ones of
ref.[2], {\bf{q}} is
reciprocal lattice vector; the crystal potential is used in the form of
$U(r)=\sum{G({\bf{q}})e^{-i{\bf{qr}}}}$ .
In numerical calculation we use the Moliere potential.  At large angles
Eqs.(3), (10)-(14) turn into the well
known equations of polarized photon emission spectra of CB theory [1].\\
We have carried out the numerical calculations at
$\varepsilon=150$ GeV for $<001>$ aligned single 
diamond crystal at temperature $T=293K$, $\vartheta_{0}$ is
the angle of the vector ${\bf{v}}_{0}$ to
this axis, and $\varphi_{0}$ is the angle of the ${\bf{v}}_{0}$ projection 
onto the plane $\{001\}$ to the plane $\{1 \bar{1}0\}$. The calculations are 
carried out for uniform distribution. On Fig.1, 2 the graduate changing of 
intensity and polarization spectra are presented when the angle $\psi$ of 
the vector ${\bf{v}}_{0}$ to the plane $\{1\bar{1}0\}$ 
$(\sin\psi=\sin \vartheta_{0}\sin \varphi_{0})$ is decreasing.  
The behavior of the unpolarized intensity coincides, in general, with
results of ref.4. The shape of the spectra is determined by the competition
of coherent effects, which are determined by high time region
$\tau\geq1/q_{z}$ in the integrals over time in (3), and magnetic 
bremsstrahlung effects, which come from low time region $\tau\ll1/q_{z}$.  
As compared with the CB results, the intensity of coherent peaks decreases, 
the peaks change their form and  move down to low frequencies $\omega$. These 
changes are partly stipulated by renormalization of electron mass in strong 
crystalline fields. 
One can also see the arising of new peaks and deeps [4], 
which also move to low $\omega$ region.
At the same time, when $\psi$ decreases, magnetic bremsstrahlung effects 
begin to contribute and bring to overall increasing of intensity. They are 
stronger displayed at higher electron energies and lower radiated frequencies, 
because the effective upper integration limit in (3),
$\tau_{0}\geq c\cdot\varepsilon\varepsilon'/m^{2}\omega$, $(c\gg1)$ ,
increases with increasing of the electron energy and with decreasing of 
the photon energy.

\begin{wrapfigure}{r}{10.cm}
\epsfig{figure=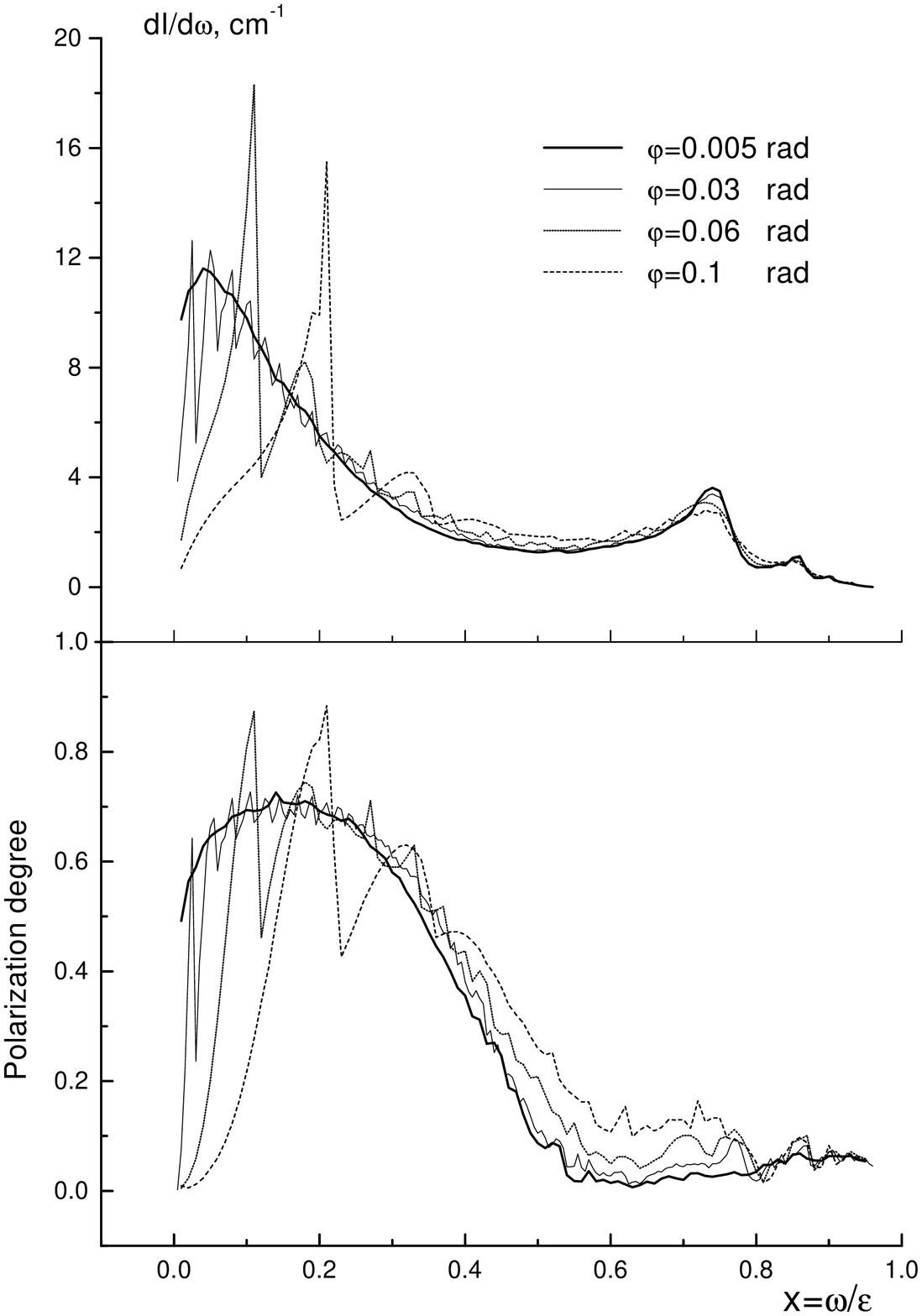,width=10.4cm}
{\small Fig.2: Changes of intensity and polarization spectra at decreasing 
of $\varphi_{0}$ at fixed $\vartheta_{0}$=0.3 mrad.}
\end{wrapfigure}
\indent In the wide range of angles the radiation has an essential linear
polarization. The circular polarization doesn't vanish but practically is 
negligible $(\xi_{2}\leq0.01)$. As
in the CB theory, the spectral distribution of the degree of polarization 
repeats, in general, the structure of the intensity distribution, i.e.
peaks of intensity and peaks of polarization are situated at the same 
frequencies. So our results differ qualitatively from the results of ref.3,  
where at the positions of the intensity maxima are situated the minima of
polarization!\\
\indent When $\varphi_{0}$ is decreasing at fixed $\vartheta_{0}$ (Fig.2) 
(transfer to purely planar orientation), 
vectors $q_{z}$ which contribute to sums in Eq.[3], form different groups,
in such a way, that each vector $q_{z}$ from the given group tends to 
the same limit at $\varphi_{0}\rightarrow0$, $q_{z}\rightarrow q^{(i)}_{z}$, 
(i=1,2,3,...). The first group of vectors which tend to zero, $q^{(i)}_{z}=0$, 
determine the magnetic bremsstrahlung contribution in the range of small 
$\omega$. This contribution brought to the large maximum of the degree of 
linear polarization (0.6-0.8) at $x=\omega/\varepsilon=0.1\div0.3$,
which remains unchanged at further decrease of $\varphi_{0}$. The 
radiation is polarized in $\{110\}$ plane $(\xi_{1}\sim 0)$.
The groups of the vectors, which tend to nonvanishing limits, 
$(q^{(i)}_{z}\neq0)$ correspond to a number of high frequency peaks. If
these peaks are situated at sufficiently large frequencies ($x\geq 0.6$), 
the magnetic bremsstrahlung effects are small enough and do not
affect seriously the formation of these peaks, so the shape and position of
these peaks can be approximately described in (non modified)
coherent theory. 
\begin{wrapfigure}{r}{10.cm}
\epsfig{figure=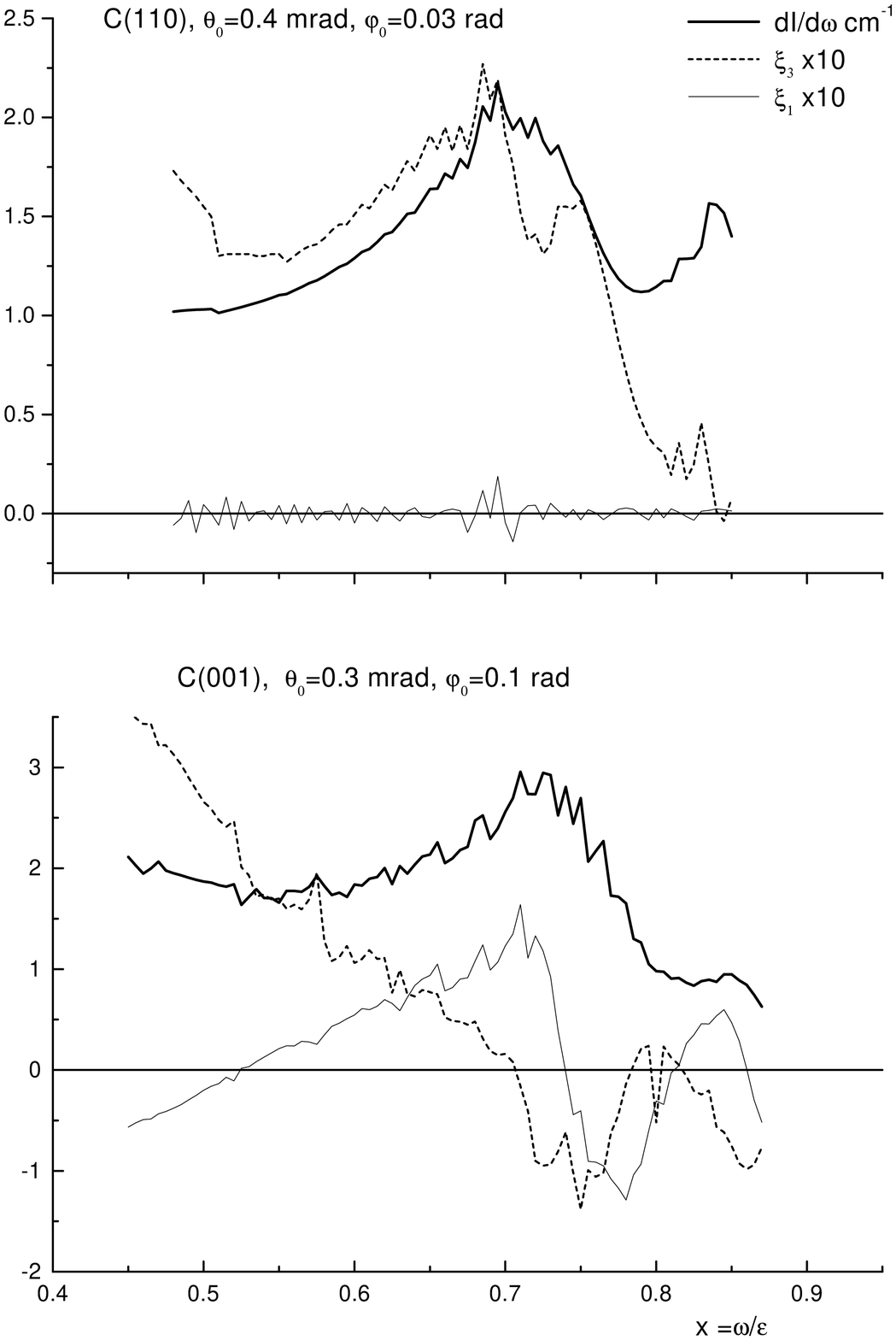,width=10.4cm}
{\small Fig.3: The spectral distribution of intensity and Stocks parameters in the
range of high-energy photon peaks for $<001>$ and $<110>$  oriented
diamond single crystal at $\varepsilon=150$Gev.}
\end{wrapfigure}
The magnitude of the peaks is determined, approximately, 
as a sum of coherent contributions of reciprocal vectors with given 
$q^{(i)}_{z}$. For the intensities of polarized photon emission, these 
contribution could be of different sign, or some contribution could vanish. 
Therefore, the polarization in the region of high $\omega$ peaks  strongly 
depends on crystal orientation and is relatively small. For instance, 
for $<001>$ aligned crystal, in the region of main high  frequency peak, 
the individual contributions practically compensate each other and degree 
of polarization is very small. For $<110>$ orientation, in the range of 
similar peak the main contributions appear to be of the same sign, and 
degree of polarization reaches 0.25 (Fig.3).

\indent Let us also note that, despite the planar orientation of crystal, 
the direction of polarization at these peaks is not immediately connected 
with the position of main plane. We have, in general, $|\xi_{1}|\sim|\xi_{3}|$. 
It is easy to understand, because according to what was explained above, 
the main plain practically do not contributions to these peaks, contributions 
come only from neighbouring planes, which are situated at    
different angles to the main plane. The "fine structure" of the spectral
distribution of intensity and Stocks parameters for different crystal
orientations in the range of high frequency peak are demonstrated in
Fig.3.\\ 
\indent High-energy photon pronounced maxima in the radiation emission have been
observed experimentally at CERN [6] in diamond and Si crystals. 
A discussion is going on as to the possibility of utilizing this effect for 
producing high-energy polarized photon beams. \\
\indent More precise calculations for different crystal orientations in 
the wide range of incident electron angles and energies we are going to 
present in subsequent papers.\\

\indent This work was supported by the ISTC grant, project A-099. One of us (N.T) 
is grateful to Prof. R.Avakian for suggestion to participate in this project.

\end{document}